\newcommand{\AJ}[3]{ #1, AJ, #2, #3}
\newcommand{\ApJ}[3]{ #1, ApJ, #2, #3}
\newcommand{\ApJS}[3]{ #1, ApJS, #2, #3}

\newcommand{\AsAp}[3]{ #1, A\&A, #2, #3}
\newcommand{\AsAs}[3]{ #1, A\&A, #2, #3}
\newcommand{\AsApS}[3]{ #1, A\&AS, #2, #3}

\newcommand{\ARAsAp}[3]{ #1, ARA\&A, #2, #3}

\newcommand{\MN}[3]{ #1, MNRAS, #2, #3}
\newcommand{\MNRAS}[3]{ #1, MNRAS, #2, #3}

\def\SAASFEE82{ 1982, Morphology and Dynamics of Galaxies, ed. L.\,Martinet,
   M.\,Mayor (Sauverny: Geneva Obs.)}

\def\IAUP{ 1987, IAU Symp. No.\ 127, Structure and Dynamics of
Elliptical Galaxies (Princeton), ed.\ T.\,de~Zeeuw (Dordrecht: Reidel)}

\def\NormalGal{ 1987, Nearly Normal Galaxies, ed. S.M.\,Faber
   (Berlin: Springer--Verlag)}

\def\ESOSTECF{{ 1989, First ESO/ST--ECF Data Analysis Workshop}, eds.\
   P.J.\,Grosb{\o}l, F.\,Murtagh, \and\ R.H.\,Warmels (Garching: ESO)}

\def\Heid1990{ 1990, Dynamics and Interactions of Galaxies, ed. R.\,Wielen
    (Heidelberg:  Springer--Verlag)}
\def\mHeid1990{ 1990. In: {\it Dynamics and Interactions of Galaxies\/}, ed.
    Wielen, R., Springer--Verlag, Berlin}

\def\REF{\par\noindent\hangindent 20pt}

\def\BGA{\begin{array}}
\def\EDA{\end{array}}
\def\BGD{\begin{description}}
\def\EDD{\end{description}}
\def\BGF{\begin{figure}}
\def\EDF{\end{figure}}
\def\BGE{\begin{equation}}
\def\EDE{\end{equation}}
\def\BGC{\begin{center}}
\def\EDC{\end{center}}
\def\BGT{\begin{tabular}}
\def\EDT{\end{tabular}}
\def\BGIT{\begin{itemize}}
\def\EDIT{\end{itemize}}
\def\BGEN{\begin{enumerate}}
\def\EDEN{\end{enumerate}}
\def\bn{\bigskip\noindent}

\def\mn{\medskip\noindent}

\def\pn{\par\noindent}
\def\sn{\smallskip\noindent}
\def\and{{\&}}

\def\Bss{\mbox{B--mag~arcsec$^{-2}$}}
\def\cf{{\it cf.\/}}

\def\dex1{\mbox{dex}}
\def\dex{\hbox{\rm dex}}

\def\eg1{{e.g.\/}}

\def\etal{{\it et al.\/}}

\def\gp{\hbox{\rlap{\hbox{.}}\raise 5.truept \hbox{{\small$\circ$}}}}
\def\gradip{\hbox{\rlap{\hbox{.}}\raise 5.truept \hbox{{\small $\circ$}}}}

\def\magcir{\ \raise-2.truept\hbox{\rlap{\hbox{$\sim$}}
\raise5.truept\hbox{$>$}\ }}
\def\me{\mbox{$\mu_e$}}

\def\micron{\mbox{$\mu$m}}
\def\mincir{\ \raise-2.truept\hbox{\rlap{\hbox{$\sim$}}
\raise5.truept\hbox{$<$}\ }}

\def\rq{\mbox{$r^{1/4}$}}

\def\secondip{\hbox{\rlap{\hbox{.}}\hbox{$''$}}}

\def\underline{}

\def\rmplane{\mbox{(\lre, \me)} \mbox{plane}}
\def\mrplane{\mbox{(\lre, \me)} \mbox{plane}}
\def\lre{\mbox{$\log\re$}}

\def\me{\mbox{$\mu_e$}}

\def\re{\mbox{$R_e$}}
\def\bright{`{\it bright\/}'}
\def\ordinary{`{\it regular\/}'}
\def\umpomeno{\vspace*{-0.15cm}}
\def\schippa{\vspace*{0.2cm}}
\def\broadsample{{\it broad--sample\/}}
\def\Md{\mbox{$M_{dust}$}}
\def\Mg{\mbox{Mg$_2$}}
\def\aq{\mbox{$a_4/a$}}
\def\Sect{\mbox{$\S$}}
\def\REF{\par\noindent\hangindent 15pt}
\def\pn{\par\noindent}
\def\sn{\smallskip\noindent}
\def\mn{\medskip\noindent}
\def\bn{\bigskip\noindent}
\def\Mmplane{\mbox{($M_B,\me$)} \mbox{plane}}
\def\rmplane{\mbox{($\lre,\me$)} \mbox{plane}}
\def\mrplane{\mbox{($\lre,\me$)} \mbox{plane}}
\def\Imed{\mbox{$\langle I \rangle_e$}}
\def\mumed{\mbox{$\langle \mu \rangle_e$}}
\def\const{\mbox{\rm const.}}
\def\bright{`{\it bright\/}'}
\def\ordinary{`{\it ordinary\/}'}

\def\me{\mbox{$\mu_e$}}

\def\gp{\hbox{\rlap{\hbox{.}}\raise 5.truept \hbox{{\small$\circ$}}}}

\def\pgr{\phantom{\Big\{}}
\def\ss{\phantom{00000000}}

\documentstyle[12pt
]{article}
\begin{document}
\baselineskip=18pt
\begin{center}
{\bf THE $(\lre,\me)$ PLANE OF HOT STELLAR SYSTEMS}\\
\bigskip\bigskip
{\it Massimo Capaccioli$^1$, Nicola Caon$^2$, Mauro D'Onofrio$^1$}\\
\bigskip
$^1$\, Department of Astronomy, University of Padova, I--35122 Padova\\
$^2$\, International School for Advanced Studies, I--34014 Trieste
\end{center}
\bigskip

\section*{\normalsize ABSTRACT}\umpomeno
Two families of hot stellar systems, named \ordinary\ and \bright, are
identified in the \rmplane\ built with a luminosity--limited sample of
ellipticals and bulges of S0s and spirals of the Virgo and Fornax clusters.
This finding, based on {\it ad hoc\/} new observations, is confirmed by a much
larger set of literature data for $\sim1500$ galaxies.
The \ordinary\ family is biparametric: $L_T\propto I_e\re^2$;
its members are fainter that $M_B\simeq-19.3$ and smaller than $\re\simeq3$
kpc (whatever $M_B$ is).
The \bright\ family is uniparametric (\me\ depends on \re\ alone) and hosts
brightest cluster members and QSO parent galaxies.

We show that the segregation in the \rmplane\ has an important counterpart in
the behavior of various physical parameters, which is markedly different for
galaxies smaller (\ordinary\ family) and larger (\bright\ family) than
$\re=3$ kpc.

\bigskip
\section*{\normalsize 1.\ \ INTRODUCTION}\umpomeno
In the last decade much effort has been put into the search for the smallest
set of independent parameters allowing a complete description of galaxy
properties.
Multivariate statistical analysis techniques have been extensively applied to
increasingly larger and better data--sets (Brosche 1973, Whitmore 1984,
Watanabe, Kodaira \and\ Okamura 1985, Djorgovski \and\ Davis 1987).
At present the existence of a two-dimensional manifold for the parameters of
hot stellar systems (E galaxies and bulges) is well established: a result
known in the literature with the pompous name of `Fundamental Plane' (FP).
The fact that the various types of galaxies populate systematically different
regions of the FP has been interpreted in terms of different formation
mechanisms (e.g. Bender, Burstein \and\ Faber 1992 =~BBF).

Among the observed or derived quantities for hot stellar systems, such as
characteristic radius and corresponding surface brightness, luminosity, mass,
mass--to--light ratio, velocity dispersion, and metallicity, there are only
two independent variables which support most of the variance.
All the other parameters can be expressed as a combination of these two
(Dressler \etal\ 1987, Faber \etal\ 1987, Kormendy \and\ Djorgovski 1989),
which are, for instance, the effective (equivalent) radius \re\ and
surface brightness $\me=\mu(\re)$, i.e. the parameters pertaining to the
isophote encompassing half the total luminosity.

In spite of the vast popularity gained by the FP, its universality has been
nonetheless questioned with regard to zero--point, slope, and thickness, and
a controversy exists concerning the differences between field and cluster
ellipticals (de Carvalho \and\ Djorgovski 1992).
A surprising fact and a matter of concern is also that opposite conclusions
about the properties of the FP are drawn from approximately the same data
set.
This is a further indication that some basic issues still call for a
clarification: two of them are the statistical completeness of the adopted
galaxy samples, and the definition of a unique methodology and of standard
procedures for data analysis.
In fact large differences can be found among total magnitudes and effective
parameters for the same galaxies studied by various authors (cf. Capaccioli,
Piotto \and\ Rampazzo 1988, and Capaccioli \etal\ 1992b).
They depend mostly on the uncertainty in setting the sky--background level
---~which plagues CCD images of large galaxies~---, on the calibration of the
photometric scale, and on the adopted extrapolation to infinity of the growth
curves, often assumed to be universal.
Moreover, much anarchy is present in the definition of the structural
parameters \me\ and \re;
they are often given as the scaling parameters of an empirical formula fitting
some light profile (e.g. the \rq\ law) rather than the parameters of the
isophote encircling half the total luminosity.

Prompted by these considerations, we have undertaken once again the
photometric mapping of large luminosity--limited samples of galaxies in
nearby clusters, adopting a methodology known as `global mapping' technique
(see Capaccioli \and\ Caon 1989).
Results for a volume--limited sample of E galaxies and of bulges of S0s and
spirals belonging to the Virgo cluster (Capaccioli, Caon \and\ D'Onofrio
1992a, hereafter Paper~I) allowed us to recognize two distinct families of
galaxies in the \mrplane, that we named \ordinary\ and \bright.
The first family consists of ellipticals and of hot galaxy components
(bulges) fainter than $M_B \simeq -19.3$ mag, with effective parameters
ranging over a large interval for the same total luminosity: \re\ varies by
$\sim 0.7\ \dex$ and \me\ by $\sim 3.5$ mag.
The interval spanned by \re\ is the same at any luminosity (down to $M_B
\simeq -12$, by adding literature data for dE and dS0 galaxies), with a sharp
upper boundary at $\re \simeq 3$ kpc.
The second is a one parameter family (\me\ depends on \re; Kormendy 1977),
consisting of galaxies with $M_B < -19.3$ and $\re >3$ kpc.
Typically they have boxy isophotes, populate cluster condensations (brightest
cluster members), and have large amplitudes of the correlation function at
small radii (Einasto \and\ Caon 1992).

In this paper we look deeper into the properties of the two families by
adding new photometric data for 25 ellipticals and S0s of the Fornax cluster,
and by searching for correlations of the structural parameters with other
observables compiled from the literature.
The paper is organized as follows.
In \Sect2 we introduce our Virgo--Fornax sample, discuss its characteristics
and completeness, and briefly comment on the data reduction.
The relation between \me\ and \re\ with our Virgo--Fornax data is shown in
\Sect3, and that built with a much larger set of literature data is presented
in \Sect4.
The correlations of the structural parameters with the other observables
are examined in \Sect5 and discussed in \Sect6.

\schippa
\section*{\normalsize 2.\ \ THE VIRGO--FORNAX SAMPLE}\umpomeno
Our photometric sample of Virgo galaxies consists of 52 Es and non--barred
S0s (Caon, Capaccioli \and\ Rampazzo 1990, Trevisani 1991), and of 54 spirals
of morphological types from Sa to Sc (D'Onofrio 1991).
According to the membership list of Binggeli, Sandage \and\ Tammann (1985),
the early--type sample is 80\% complete to the total apparent magnitude $B_T
= 14$, i.e. to the absolute magnitude $M_B = -17.3$, if a distance modulus of
$(m-M)_0 = 31.3$ is adopted after Capaccioli \etal\ (1990b).
The spiral sample is also 80\% complete to $B_T = 13.5$.
In both cases the missing objects lie mostly at the cluster outskirts, not
covered by our large-field plates.
As for the Fornax cluster, CCD and plate material has been collected for 31
E and non--barred S0 galaxies (Caon, Capaccioli \and\ D'Onofrio 1992).
With respect to the membership list of Ferguson (1989), this sample is
100\% complete to $B_T = 15$.
Here we present data for 25 galaxies reduced so far.

Since our strategy requires an accurate surface photometry covering the
largest possible range in surface brightness, we have adopted  the
so--called `global mapping' technique (Capaccioli \and\ Caon 1989).
Combining unsaturated CCD images with large--field (Schmidt) deep
photographs, it allows a mapping of the light distribution of galaxies
from the center down to $\mu_B = 28$ \Bss, with a relative accuracy better
than 0.1 mag from outside the seeing--convolved core out to $\mu_B \sim
26$.
A key feature of this technique is the sky--background subtraction in the
CCD images.
The CCD blank--sky level $\mu_s$ is determined by requiring that the sky
subtracted CCD light profile matches, in the unsaturated range, the
corresponding profile extracted from the photographic image (where $\mu_s$
can be measured with a precision better than 0.5\%).
Note that the error on $\mu_s$(CCD), usually $< 2\%$, is not better
than for other common methods of measuring the sky--background directly
on the CCD frame (at least for objects not completely filling the frame).
What the `global mapping' technique provides is the possibility of using the
photographic profile from where the CCD light profile becomes unreliable
because of the uncertainty on the blank--sky level.

Geometrical and photometric parameters for the set of early--type galaxies are
those of two--dimensional photometric models built coupling the light profiles
along the principal axes with the ellipticity and position angle profiles (see
Caon \etal\ 1990, for details which include the extrapolation technique to
compute the total magnitude and, in turn, the effective isophote parameters).
For spirals a standard bulge--disk decomposition procedure has been followed,
which gave results for 35 objects (D'Onofrio 1991).
We resorted to both the interactive and the least--square fitting technique
extensively used elsewhere (e.g. Schombert \and\ Bothun 1987).
The main axis light profiles were decomposed into the sum of a \rq\ law
(de Vaucouleurs 1948) for the bulge and an exponential law (Freeman 1970)
for the disk respectively, rejecting solutions which did not appear
plausible.

\schippa
\section*{\normalsize 3.\ \ THE ($\lre,\me$) PLANE FOR THE VIRGO--FORNAX
SAMPLE}\umpomeno
Figure~1 shows the distribution of the representative points for our 87
Virgo and 25 Fornax ellipticals and bulges in the \mrplane.
We can easily identify two groups.
The first one, that we call \ordinary\ family, appears confined within a
strip bounded by lines which are of constant luminosity for galaxies with
homologous light--distributions:
\BGE
\label{0}
L_T = s I_e R_e^2
\EDE
$s$ is a `structural parameter' whose value depends on the shape of the
galaxy light profiles;
for a Sersic (1968) law
\BGE
I(r) = I_0\, \dex\Big(-r^{1/n}\Big)
\label{Sersic}
\EDE
numerical calculations show that, in the range $0.6 < n < 10$, the
structural parameter is very accurately approximated by the relation:
\BGE
\log s = 0.46 \log n + 1.08
\EDE
The \bright\ family consist of 12 galaxies, the brightest of the sample.
Well isolated from the others, they are characterized by large effective
radii ($\langle \re \rangle = 10$ kpc) and relatively low surface brightness
($\langle \me \rangle = 23.7$ \Bss).
This group, together with the brightest members of the \ordinary\ group,
fits the uniparametric relation reported by Hamabe \and\ Kormendy (1987 =
HK):
$ \me = 2.94 \log \re +20.75$,
with zero point tuned to our Virgo cluster distance.

\begin{figure}[t]
\vspace{13.5truecm}
{\caption[]{
Sample of 87 Virgo galaxies (ellipticals and bulges of S0s and spirals)
and of 25 Fornax (Es and S0s), showing some sort of bimodal distribution
in the \mrplane.
The adopted distance modulus is 31.3 mag both for Virgo and Fornax
clusters.
The upright solid line $\lre = 0.45$ is an estimate of the upper
boundary to the effective radius for the so--called \ordinary\ family (see
text).
The diagonal solid line marks the locus of constant luminosity $M_B =
-19.3$ for homologous galaxies.
The dashed line is the HK relation, holding for the brightest group.
The drift caused by an error on the total magnitude $B_T$ is indicated by the
dotted line (distance between two ticks corresponds to $\delta B_T=0.1$ mag).
Note that the discontinuity in \lre\ between the two families might not be
real, but just a spurious consequence of the poor statistics or a peculiarity
of the local environment.}}
\end{figure}

\noindent
A least--square fit for the early--type members of the \ordinary\ family (E
and S0 galaxies of both Virgo and Fornax with $M_B > -19.3$) yields:
\BGE
M_B = -5.66(\pm0.07)\log R_e + 1.02(\pm0.01)\mu_e -40.2(\pm0.3)
\label{1}
\EDE
{}From the Virial Theorem:
\BGE
{L \over R} \left({M \over L}\right) \propto \sigma^2
\label{2}
\EDE
and from our data:
\BGE
L\propto I_e^{1.02 \pm 0.01}\, R^{2.26 \pm 0.03}
\label{3}
\EDE
By combining eqs.~\ref{2} and \ref{3}:
\BGE
L \propto I_e^{-0.81 \pm 0.11}\,\sigma^{3.58 \pm 0.13}\,\left({M \over L}
\right)^{-1.79 \pm 0.05}
\label{4}
\EDE
which compares quite well with Djorgovski \and\ Davis' (1987) FP:
\BGE
L \propto \Imed^{-0.86} \sigma^{3.45}
\label{5}
\EDE
if the mass--to--light ratio $M/L$ is marginally dependent on $L$.
(\Imed\ is the mean surface brightness within the effective isophote).
Equation~\ref{1} changes into:
\BGE
M_B = -5.40(\pm0.04)\log R_e + 1.00(\pm0.01)\mu_e -39.8(\pm0.3)
\label{1bis}
\EDE
if the \bright\ family is added to the \ordinary\ one, with which eq.~\ref{4}
becomes:
\BGE
L \propto I_e^{-0.86 \pm 0.06}\,\sigma^{3.72 \pm 0.08}\,\left({M \over L}
\right)^{-1.86 \pm 0.03}
\label{4bis}
\EDE

\noindent
The coefficient of \lre\ in eq.~\ref{1bis} differs from the expectation for an
homologous family (eq.~\ref{0} with $s= \mbox{const.}$).
Assuming that the structural parameter $s$ depends on the effective radius:
$s = u\times R_e^\gamma$, where $u$ is now a universal constant, we obtain
$\gamma=0.16$, which means that $s$ increases, though slowly, with \re.
This corresponds to a progressive shallowing of the light profile as \re\
stretches, which is indeed observed.
In fact, fitting the luminosity profiles of our galaxies with a Sersic law,
we find (Fig.~3{\it k\/}) a clear trend of the exponent $n$ with \re:
\BGE
\log n = 0.51(\pm 0.04) \log\re + 0.30(\pm 0.03)
\EDE

\noindent
{}From Fig.~1 we see that objects of equal luminosity span the ranges
$\Delta\me \sim 3.5$ mag and $\Delta\lre \sim 0.7$.
This spread is unlikely due to projection effects.
For instance, the surface brightness of an oblate spheroid at face--on
view is $2.5\log(b/a)$ fainter than at the view angle inducing the
apparent axis ratio $b/a$ of the isophotes;
therefore, going from $\varepsilon=1-b/a=0$ to $\varepsilon=0.6$, \me\
varies at most by 1 mag.
Since photometric errors are estimated to be $\delta\lre \simeq 0.08$
and $\delta\me \simeq 0.4$ (Capaccioli \and\ Caon 1991), this implies
that most of the dispersion in \me\ at the same luminosity is intrinsic,
in spite of the fact that galaxies with equal $M_B$ tend to have the same
structural parameter $s$ (or the same Sersic index $n$).
\begin{figure}[t]
\noindent
{\caption[]{
{\it Panel a\/})\ \ The \mrplane\ for the \broadsample\ of more than 1500
galaxies. Open circles are E and S0 galaxies of the Virgo--Fornax sample
measured by the `global mapping' technique.
Crosses represent the bulges of spirals from Kent (1985) and D'Onofrio (1991).
The small filled dots are data from Schneider \etal\ (1983), Thomsen \and\
Frandsen (1983), Malumuth \and\ Kirshner (1985), Hoessel \and\ Schneider
(1985), Michard (1985), Schombert (1987), Capaccioli \etal\ (1988), and
J{\o}rgensen \etal\ (1992).
Starred symbols are galaxies hosting a QSO and Seyferts from Malkan (1984) and
Malkan \etal\ (1984).
Open triangles are BC data.
Open squares are galaxies from the Fornax survey by DPCDK and Irwin \etal\
(1990); uncertain data have been disregarded.
Compact dwarf ellipticals (Bender \and\ Nieto 1990) are indicated as filled
diamonds.
The heavy solid lines correspond to $\re = 3$ kpc and to $M_B=-19.3$ mag.
The long--dashed line is the BC relation (eq.~\ref{BCeq}).
The dashed line is the HK relation.\\
{\it Panel b\/})\ \ The same data as in panel {\it a\/}), plotted in the
\Mmplane\ with the same coding for symbols of representative points and
for lines. The latter have been derived from those in panel {\it a\/}) under
the assumption of homology. }}
\end{figure}

\schippa
\section*{\normalsize 4.\ \ THE ($\lre,\me$) PLANE FOR THE
BROAD--SAMPLE}\umpomeno
In order to strengthen the above findings, we have collected literature
values of the effective parameters for as many as $\sim1400$ galaxies
(hereafter referred to as \broadsample), spanning a large range of
luminosities and morphological types.
Capaccioli \etal\ (1992b) give full account for all sources (and for their
registration to the same distance scale and color band).
We have added here the data--sets on dwarf galaxies by Davies \etal\ (1988 =
DPCDK) and by Irwin \etal\ (1990), and on compact dwarf Es from the
compilation by Bender \and\ Nieto (1990).

The \broadsample\ is plotted in Fig.~2a.
The solid lines mark the boundaries for the region of the \ordinary\ family:
$\re < 3$ kpc, $M_B > -19.3$ mag.
The short--dashed line represents the HK relation for the \bright\ family.
The marked gap between the two families, which was present in our data
(Fig.~1), has now disappeared.
At the moment we are unable to establish whether this gap is real (and
therefore it is wiped out, in Fig.~2a, by the heterogeneity of the
\broadsample\ data) or whether it is a consequence of the poor statistics of
our Virgo--Fornax sample.
We shall note that the first hypothesis rests on the circumstance that the
error vector is roughly aligned with the HK relation (cf. Capaccioli \etal\
1992b); thus, large errors tend to fill the gap, if any.
The long--dashed line in the figure reproduces the fit by Binggeli \and\
Cameron (1992 = BC) to their data--set of early--type dwarf galaxies (see
below).

The \broadsample\ has also been plotted in the \Mmplane\ of Fig.~2b
in order to discuss the results of BC and DPCDK.
BC suggest that dwarf and faint ellipticals and S0s (our \ordinary\ family)
follow the surface brightness--luminosity relation (reduced to our Virgo
distance):
\BGE
\mumed \simeq 0.75 M_B + 34.98
\label{BCeq}
\EDE
\vfill
\newpage
\noindent
with a considerable scatter (0.8 mag at 1 sigma level), mostly cosmic.
Here $\mumed= -2.5\log\Imed+\const$;
for an homologous family $\mumed = \me - k$,
where $k=1.15\log n + 0.70$ for a Sersic formula.
The slope of eq.~\ref{BCeq} is not far from unity, that is from the condition
$\re = \mbox{const.}$, and indeed, the line at $\re = 3$ kpc corresponds
almost perfectly to the lower envelope of the dwarf distribution.

We fully agree with BC that the low--luminosity Es ($M_B>-19.3$) do not share
the $M_B$--\mumed\ relation defined by the giants.
Fig.~2b suggests that low--luminosity Es simply extend the dwarf
distribution toward brighter $M_B$ and $\mumed$.
Similarly, we are not in the position of establishing whether the compact
dwarf ellipticals (prototype M32), which populate the upper left part of
\mrplane\ (see Fig.~2a), are a third family or ---~less likely, though~---
an extension of the distribution of the \ordinary\ family to very small
objects which, due to their compactness and proximity to giant galaxies,
are still observable/observed.

In this context we shall comment on the claim by DPCDK that the representative
points of their Fornax cluster dwarfs are spread in the effective parameters
plane (open squares in Figs.~2).
Part of these data are likely flawed by insufficient angular resolution (cf.
Ferguson \and\ Sandage 1988), so that the exponential fitting of luminosity
profiles can give unreliable results.
Moreover, some of the DPCDK galaxies are suspected to be background
objects, a fact which adds confusion to the ($M_B,\me$) distribution.

\schippa
\section*{\normalsize 5.\ \ CORRELATIONS BETWEEN STRUCTURAL AND PHYSICAL
PARAMETERS}\umpomeno
Does the the existence of two families in the \mrplane\ bear any physical
meaning\,?
In an attempt to answer this basic question, we have compiled from the
literature a catalogue of observables for the early--type galaxies of our
Virgo--Fornax sample.
We aim at testing if the line $\re = 3$ kpc dividing the two families in
the effective parameters plane is also a `watershed' for other physical
properties.
In particular we have looked at the 21\,cm line emission, the radio
continuum flux at 6\,cm, the CO mappings at 2.6\,mm, the IRAS fluxes from
12 to 100\,\micron, the UV luminosities, the X--ray data in the 0.2--4 keV
band, and the masses for the interstellar medium components.
We also considered other photometric and spectroscopic quantities such as
color gradient, metallicity index, isophotal shape parameters, maximum
ellipticity, central velocity dispersion, and anisotropy parameter.

\begin{table}[tb]
\begin{center}
{Table 1: Data from the literature}\\[12pt]
\begin{tabular}{lll}
\hline\hline\\[-7pt]
Parameter & {\ \ \ \ \ Data Source} & {\ \ No. of detections} \\
          &                         & {(No. of upper limits)} \\[3pt]
\hline\\[-10pt]
$100*(\aq)_{peak}$          &\pgr Caon \etal\ 1992            & \ss 77
\phantom{(99)} \\
$\varepsilon_{max}$         &\pgr  Caon \etal\ 1992           & \ss 77
\phantom{(99)} \\
\Mg\ index                  &\pgr  Davies \etal\ 1987         & \ss 30
\phantom{(99)} \\
$(V/\sigma)^*$              &\pgr  Bender \etal\ 1992         & \ss 18
\phantom{(99)} \\
                            &\pgr  Knapp \etal\ 1985 and      & \\[-10pt]
HI mass                     &\Big\{                           &
\ss \phantom{991}
(40) \\[-10pt]
                            &\pgr  Wardle \and\ Knapp 1986    & \\
Dust mass                   &\pgr  Roberts \etal\ 1991        & \ss 18
\phantom{1}(23) \\
$L_X$                       &\pgr  Fabbiano \etal\ 1992       & \ss 16
\phantom{1}(10) \\
$L_{6\,cm}$                 &\pgr  Roberts \etal\ 1991        & \ss 15
\phantom{1}(23) \\
                            &\pgr  Vader \etal\ 1988          & \ss 16
\phantom{(99)}\\[-10pt]
Color gradients             &\Big\{                           & \\[-10pt]
                            &\pgr  Peletier \etal\ 1990       & \ss 10
\phantom{(99)} \\[4pt]
\hline
\end{tabular}
\end{center}
\end{table}

\noindent
In Table~1 we list those parameters which show some sort of dependence on the
effective radius \re\ (Figs.~3).
The sources of the data and the numbers of galaxies used are in cols.~2 and 3
respectively.
Unfortunately the intersections of our Virgo--Fornax photometric
data--base with the lists of UV magnitudes (Longo, Capaccioli \and\ Ceriello
1991), CO abundances (Roberts \etal\ 1991), and SNe events (Barbon,
Cappellaro \and\ Turatto 1989) are almost empty.
No significant correlations exist with dustiness alone (data from van der Bergh
\and\ Pierce 1990), with HI flux alone (from Roberts \etal\ 1991), with IRAS
fluxes at 12, 25, 60 and 100 $\micron$ (from Knapp \etal\ 1989), with
$L_{IR}/L_B$ (from Bally \and\ Thronson 1989), and with the residuals from the
Faber--Jackson relation.
Positive correlations are shown in the panels from {\it a\/}) to {\it j\/}) of
Fig.~3.

\noindent
Fig.~3{\it a\/}: galaxies with a large \re\ tend to have boxy--shaped or
elliptical isophotes.
Note that this correlation is degraded by the projection effects which
influence both \aq\ and \re.

\noindent
Fig.~3{\it b\/}: strong ellipticities are found only in the \ordinary\
family.
Should this result be confirmed by a larger sample, it would imply that the
light distribution in \bright\ galaxies is close to spherically symmetric
(cf. Capaccioli, Caon \and\ Rampazzo 1990a).

\noindent
Fig.~3{\it c\/}: the \Mg\ index is more dispersed for the \ordinary\ galaxies.
This is expected in view of the result of Fig.~3{\it a\/} and of the
finding by Longo \etal\ (1989) of some dependence of the \Mg\ index on \aq.

\noindent
Fig.~3{\it d\/}: \bright\ galaxies have a larger dispersion of the
anisotropy parameter $(V/\sigma)^*$ (Davies \etal\ 1983).
According to BBF this parameter is a crude measure of the ratio of baryonic
mass in (cold) gas to mass in stars at the time when the last major merger
occurred.

\noindent
Fig.~3{\it e\/}: \bright\ galaxies seem to possess a low HI--mass per unit of
B--light.
Note however that the HI data are just upper limits.

\noindent
Fig.~3{\it f\/}:
The mass of cold dust, computed by Roberts \etal\ (1991) from the 60 and
100 \micron\ IRAS fluxes, is related here to the total luminosity in the
B--band.
In this sample of early--type galaxies the $\Md/L_B$ ratio is very low for the
\bright\ objects.

\noindent
Figs.~3{\it g\/} and {\it h\/}:
Radio and X--ray fluxes correlate with \re\ (as they do with \aq; cf.
Bender \etal\ 1989).
Bright E and S0 galaxies are known to be dominated by the emission from a hot
interstellar medium and to be powerful X--ray emitters.
The radio luminosity is also large for these objects.

\noindent
Fig.~3{\it i\/}:
The color gradients of \bright\ galaxies present a very small dispersion.
Vader \etal\ (1988) found a dependence of color gradient on absolute
magnitude and on rotational velocity, a fact which led them suspect the
existence of two families of early--type galaxies.
Color gradients provide useful hints to galaxy formation theories,
since dissipation and merging produce different effects on galaxy colors
(\cf\ White 1979 and Carlberg 1984).
However, a recent study by Peletier \etal\ (1990) does not confirm the large
scatter in the color gradients for galaxies of different luminosities.

\noindent
Fig.~3{\it j\/}:
The luminosity density, now plotted versus $M_B$ to help comparison with
Djorgovski (1992), shows opposite trends for the two families.
(Here the vertical line is at $M_B=-19.3$ mag, the luminosity boundary
between \ordinary\ and \bright\ galaxies).
This behavior is in marked contrast to that claimed by Djorgovski (1992);
his straight--line fitting to the distribution ($\rho_L \propto L^{-0.9}$;
dotted line) is due to a lack of data for fainter galaxies.
Several processes could in principle modify the densities of proto-ellipticals:
dissipative collapse increases the density at a fixed mass, dissipationless
merging increases the mass, but decreases the density, dissipative merging
increases both the mass and density, and galactic winds by SNe decrease both.
A non--trivial mechanism should regulate the maximum density achievable by the
stellar systems ($\sim 1$ L$_{\odot}$~pc$^{-3}$) and the maximum effective
radius ($\re = 3$ kpc).

An interesting correlation is found between \re\ and the exponent $n$ of the
Sersic law (eq.~\ref{Sersic}) best-fitting the observed light profiles.
$n$ increases (possibly linearly) with $\lre$ (Fig.~3{\it k\/} and
eq.~11).
This fact seems important since it is generally believed that the parameters
related to the shape of the isophotes and to the light distribution are poorly
correlated with the variables of the FP.
As an example, in Fig.~3{\it l\/} we show that the average light profile
for the 10 brightest galaxies of our sample is very well represented by a
Sersic law with $n=6.5$. (Seeing convolution has been computed for average
conditions $FWHM=1\secondip6$ under the assumption of circular symmetry.)
This behavior reflects the homogeneity of the \bright\ family and calls for a
unique formation mechanism.

\schippa
\section*{\normalsize 6.\ \ DISCUSSION}\umpomeno
The existence of two families of galaxies in the \mrplane\ is clearly seen in
Figs.~1 and 2{\it a},{\it b}, in spite of the heterogeneity of the literature
data.
Remarkable features are that galaxies with luminosities differing by as
much as $3\,\dex$ nonetheless share the same range of \re\, and that, whatever
the luminosity is, the \ordinary\ galaxies do not grow larger in size than
$\re \simeq 3$ kpc.
The break in the distribution of the structural parameters calling for the
existence of two main families of galaxies occurs at $M_B \simeq -19.3$.

We have studied here the correlations of the effective radius \re\ with
different physical observables, keeping in mind that \re\ is not affected by
the same problems as the \aq\ parameter proposed by Bender \etal\ (1989).
(The limits of \aq\ have been analyzed by Rix \and\ White 1990, and
Stiavelli \etal\ 1991, who pointed out that it depends strongly on the
inclination and on the intrinsic shape of the galaxy as well as on the
specific view angle.)
A clear separation in the properties of the two families is evident for the
many physical parameters presented in the panels of Fig.~3.

The results of Figs.~3 should not be interpreted in a strictly statistical
sense, because the completeness of the data sample is poor and we have not
applied any test to determine the confidence level of the observed
correlations, or to establish if the data can be drawn from two different
populations.
Thus we shall refrain from expanding further on the speculations on
the nature of the two families.
None of the evidence presented here is in contrast with the view (Capaccioli
\etal\ 1992a,b) that the \ordinary\ family is a genetic variety while the
\bright\ family is the result of some sort of environmental evolution.
On the contrary, they all seem to reinforce this picture which, however,
has some other difficulties of its own; see, for instance, the discussion
in Wielen (1990) and Barbuy \and\ Renzini (1992).

An interesting fact to note is that, while ordinary Es ($M_B<-18.0$) and bulges
form a discontinues sequence with respect to dwarfs in the core--parameters
diagram ($M_B,\mu_0$) (see Kormendy 1985, and BC), such behavior is not
present in the \mrplane.
Here these objects populate a continuous distribution, interrupted at $M_B
\simeq -19.3$, i.e. at the breakpoint between our two families.
Why global parameters provide a different behavior than core parameters is
still unclear.

\schippa

\section*{\normalsize REFERENCES}\umpomeno\umpomeno

\REF Bally, J., \and\ Thronson, H.A. \AJ{1989}{97}{69}

\REF Barbon, R., Cappellaro, E., \and\ Turatto, M. \AsApS{1989}{81}{421}

\REF Barbuy, B., \and\ Renzini, A. (eds.), 1992, IAU Symp. 149, The Stellar
Populations of Galaxies, (Dordrecht: Kluwer)

\REF Bender, R., \and\ M{\"o}llenhoff, C. \AsAp{1987}{177}{71}

\REF Bender, R., \and\ Nieto, J.-L. \AsAp{1990}{239}{97}

\REF Bender, R., Burstein, D., \and\ Faber, S.M. \ 1992, preprint (=~BBF)

\REF Bender, R., Surma, P., D{\"o}bereiner, S., M{\"o}llenhoff, C., \and\
Madejsky, R. \AsAs{1989}{217}{35} (=~BSDMM)

\REF Binggeli, B., \and\ Cameron, L.M. \ 1992, A\&AS, in press (=~BC)

\REF Binggeli, B., Sandage, A., \and\ Tammann, G. \AJ{1985}{90}{1681}

\REF Brosche, P. \AsAp{1973}{23}{259}

\REF Caon, N. Capaccioli, M., \and\ Rampazzo, R. \AsApS{1990}{86}{429}

\REF Caon, N. Capaccioli, M., \and\ D'Onofrio, M. \ 1992, in preparation

\REF Capaccioli, M., \and\ Caon, N. \ \ESOSTECF, p. 107

\REF Capaccioli, M., \and\ Caon, N. \MN{1991}{248}{523}

\REF Capaccioli, M., Caon, N., \and\ D'Onofrio, M. 1992a, MNRAS, in
press (=~Paper~I)

\REF Capaccioli, M., Caon, N., D'Onofrio, M., \and\ Trevisani, S. \ 1992b, in
New Results on Standard Candles, ed. F. Caputo, Mem. SAIt, 63, p. 509

\REF Capaccioli, M., Caon, N., \and\ Rampazzo, R. \MN{1990a}{242}{24p}

\REF Capaccioli, M., Cappellaro, E., Della Valle, M., D'Onofrio, M., Rosino,
L., \and\ Turatto, M. \ApJ{1990b}{350}{110}

\REF Capaccioli, M., Piotto, G., \and\ Rampazzo, R. \AJ{1988}{96}{497}

\REF Carlberg, R.G. \ApJ{1984}{286}{403}

\REF Davies, R.L., Burstein, D. Dressler, A., Faber, S.M., Lynden-Bell, D.,
Terlevich, R.J., \and\ Wegner, G. \ApJS{1987}{64}{581}

\REF Davies, R.L., Efstathiou, G., Fall, S.M., Illingworth, G., \and\
Schechter, P.L. \ApJ{1983}{266}{41}

\REF Davies, J.I., Phillipps, S., Cawson, M.G.M., Disney, M.J., \and\
Kibblewhite E.J. \MN{1988}{232}{239} (=~DPCDK)

\REF de Carvalho, R.R., \and\ Djorgovski, S. \ApJ{1992}{389}{L49}

\REF de Vaucouleurs, G. \ 1948, Ann. Ap., 11, 247

\REF Djorgovski, S. 1992, in {\it Cosmology and Large--Scale Structure in the
Universe\/}, ed. R.R. de Carvalho (ASP Conf. Ser.), p. 19

\REF Djorgovski, S., \and\ Davis, M. \ApJ{1987}{313}{59}

\REF D'Onofrio, M. \ 1991, Ph.D. Thesis, I.S.A.S., Trieste

\REF Dressler, A., Lynden--Bell, D., Burstein, D., Davies, R.L., Faber, S.M.,
Terlevich, R.J., \and\ Wegner, G. \ApJ{1987}{313}{42}

\REF Einasto, M., \and\ Caon, N. \ 1992, MNRAS, submitted

\REF Fabbiano, G., Kim, D.-W., \and\ Trinchieri, G. \ApJS{1992}{80}{531}

\REF Faber, S.M., Dressler, A., Davies, R.L., Burstein, D., Linden--Bell,
D., Terlevich, R., \and\ Wegner, G. \NormalGal, p. 175

\REF Ferguson, H.C. \AJ{1989}{98}{367}

\REF Ferguson, H.C., \and\ Sandage, A. \AJ{1988}{96}{1520}

\REF Freeman, K.C. \ApJ{1970}{160}{811}

\REF Hamabe, M., \and\ Kormendy, J. \IAUP, p. 379 (=~HK)

\REF Hoessel, J.G., \and\ Schneider, D.P. \AJ{1985}{90}{1468}

\REF Irwin, M.J., Davies, J.I., Disney, M.J., \and\ Phillipps, S.
\MN{1990}{245}{289}

\REF J{\o}rgensen, I., Franx, M., \and\ Kj{\ae}rgaard, P. \ 1992,
preprint

\REF Kent, S.M. \ApJS{1985}{59}{115}

\REF Knapp, G.R., Guhathakurta, P., Kim, D.-W., \and\ Jura, M.
\ApJS{1989}{70}{329}

\REF Knapp, G.R., Turner, E.L., \and\ Cunnifee, P.E. \AJ{1985}{90}{454}

\REF Kormendy, J. \ApJ{1977}{218}{333}

\REF Kormendy, J. \ApJ{1985}{292}{L9}

\REF Kormendy, J., \and\ Djorgovski, S. \ARAsAp{1989}{27}{235}

\REF Longo, G., Capaccioli, M., \and\ Ceriello, A. \AsApS{1991}{90}{370}

\REF Longo, G., Capaccioli, M., Bender, R., \and\ Busarello, G.
\AsAp{1989}{225}{L17}

\REF Malkan, M.A. \ApJ{1984}{287}{555}

\REF Malkan, M.A., Margon, B., \and\ Chanan, G.A. \ApJ{1984}{280}{66}

\REF Malumuth, E.M., \and\ Kirshner, R.P. \ApJ{1985}{291}{8}

\REF Michard, R. \AsApS{1985}{59}{205}

\REF Peletier, R., Davies, R.L., Illingworth, G.D., Davis, L.E., \and\
Cawson, M. \AJ{1990}{100}{1091}

\REF Rix, H.-W., \and\ White, S.D.M. \ApJ{1990}{362}{52}

\REF Roberts, M.S., Hogg, D.E., Bregman, J.N., Forman, W.R., \and\ Jones, C.
\ApJS{1991}{75}{751}

\REF Schneider, D.P., Gunn, J.E., \and\ Hoessel, J.G. \ApJ{1983}{268}{476}

\REF Schombert, J.M. \ApJS{1987}{64}{643}

\REF Schombert, J.M., \and\ Bothun, G.D. \AJ{1987}{92}{60}

\REF Sersic J.-L. \ 1968, Atlas de Galaxias Australes
(Cordoba: Observatorio Astronomico).

\REF Stiavelli, M., Londrillo, P., \and\ Messina, A. \MN{1991}{251}{57p}

\REF Thomsen, B., \and\ Frandsen, S. \AJ{1983}{88}{789}

\REF Trevisani, S. \ 1991, Laureato Thesis, Univ. of Padova

\REF Vader, J.P., Vigroux, L., Lachieze-Rey, M., \and\ Souviron, J.
\AsAp{1988}{203}{217}

\REF van der Bergh, S., \and\ Pierce, M. \ApJ{1990}{364}{444}

\REF Wardle, M., \and\ Knapp, G.R. \AJ{1986}{91}{23}

\REF Watanabe, M., Kodaira, K., \and\ Okamura, S. \ApJ{1985}{292}{72}

\REF White, S. \MNRAS{1979}{189}{831}

\REF Whitmore, B.C. \ApJ{1984}{278}{61}

\REF Wielen, R. (ed.), 1990, Dynamics and Interaction of Galaxies, (Springer
Verlag: Heidelberg)

\end{document}